\documentclass[12pt]{article}
\usepackage{a4wide}
\usepackage{latexsym, amssymb,amsbsy}
\usepackage{here}
\newif\ifpdf
        \ifx\pdfoutput\undefined
        \pdffalse 
        \else
        \pdfoutput=1 
        \pdftrue
        \fi
\ifpdf
        \usepackage[pdftex]{graphicx}
        \DeclareGraphicsExtensions{.pdf, .jpg}
\else
        \usepackage[dvips]{graphicx}
                \DeclareGraphicsExtensions{.eps, .jpg}
\fi
\def\vec#1{\boldsymbol{#1}}
\begin{document}
\title{\Large\bf Safe domain and elementary geometry}
\author{Jean-Marc~Richard\\[2pt]
\small\sl Laboratoire de Physique Subatomique et Cosmologie, \\[-3pt]
\small\sl  Universit\'e Joseph Fourier-- CNRS-IN2P3, \\[-3pt]
\small\sl 53, avenue des Martyrs, 38026 Grenoble cedex, France}
\date{\small\textsf{\today}\normalsize}
\maketitle
\begin{abstract}
A classical problem of mechanics involves a projectile fired from a given point
with a given velocity whose direction is varied. This results in a family of
trajectories whose envelope defines the border of a ``safe'' domain. In the
simple cases of  a constant force,
 harmonic potential, and Kepler or Coulomb motion, the trajectories are conic
curves whose envelope in a plane is another conic section which can be derived
either by simple calculus or by geometrical considerations. The case of harmonic
forces reveals a subtle  property of the maximal sum of distances within an
ellipse.
\end{abstract}
%
\section{Introduction}\label{se:Intro}
A classical problem of classroom mechanics and military academies  is the
border of the so-called safe domain. A projectile is set off from a point $O$
with an initial velocity $\vec{v}_0$ whose modulus is fixed by the intrinsic
properties of the gun, while its direction can be varied arbitrarily. 
Its is
well-known that, in absence of air friction, each trajectory is a
parabola, and that, in any vertical plane containing $O$, the envelope of all
trajectories is another parabola, which separates  the points which can be shot
from those which are out of reach. This will be shortly reviewed in
Sec.~\ref{Constant}. Amazingly, the problem of this envelope parabola can be
addressed, and solved, in terms of elementary
geometrical properties.

In elementary mechanics, there are similar problems, that can be solved
explicitly and lead to families of conic trajectories whose envelope is also a
conic section. Examples are the motion in a Kepler or Coulomb field, or in an
harmonic potential. This will be the subject of Secs.~\ref{Kepler} and
\ref{Harmo}.

It is intriguing, that the property 
of ellipses unveiled by the case of the harmonic potential is not very well
known (following an e-mail survey around some mathematician colleagues), and is
not easily proved by simple geometrical reasoning. It turns out, actually, that
the mechanics problem provides one of the simplest sets of equations leading to
 the desired proof. 

This contrasts with the problem of Kepler ellipses. Here, the purely
geometrical proof is astonishingly simple, and overcomes in efficiency and
elegance the proof that can be written down by elementary calculus.
It is hoped that students will be encouraged  to carry out the solution to
these problems from both a geometrical view point and a sober handling of the
basic equations.

Kepler motion and other classical problems of elementary mechanics have been
treated very elegantly in several textbooks and articles, a fraction of which
insist convincingly on the geometrical aspects. It is impossible to quote
here all relevant pieces of the literature. Some recent articles 
 \cite{AJP-EJP} allow one to trace back
many previous contributions.

In particular, the problem of the safe domain in a constant field is well
treated in Ref.~\cite{Donnelly}, where the point of view of successive
trajectories of varied initial angle and the point of view of simultaneous
firing in all directions are both considered. The case of Coulomb or Kepler
motion is treated in some detail by
French \cite{French}, with references
to earlier work by Macklin, who used geometric methods. The case of Rutherford
scattering starting from infinite distance can be found, e.g., in a paper by
Warner and Huttar \cite{Warner}, and in Ref.~\cite{French}, while the case of
finite initial distance is discussed in a paper by Samengo
and Barrachina \cite{Barrachina}. Hence, we shall include in our discussion the
cases of constant force and inverse squared-distance force only for the sake of
completeness. The envelope of ellipses in a harmonic potential is also treated
in Ref.~\cite{French}, but with standard envelope calculus. The geometric
approach presented here is new, at least to our knowledge.
\section{Constant force}\label{Constant}
Let us assume a constant force $\vec{f}$ whose direction is chosen as the
vertical axis. This can be realized as the gravitational field in ballistics,
or an electric field acting on non-relativistic charged particles. It is
sufficient to consider a meridian plane $Oxz$. If $\alpha$ denotes the angle
of the initial velocity $\vec{v}_{0}$ with respect to the $x$ axis, then the
motion of a projectile fired from the origin $O$ is
\begin{equation}\label{C:eq:para1}
x=v_0 t\cos\alpha~,\quad
y=v_0 t\sin\alpha+{f t^2\over 2m}~.
\end{equation}
\subsection{A family of parabolas}
\begin{figure}[!!ht]
\begin{center}
\includegraphics[width=.5\textwidth]{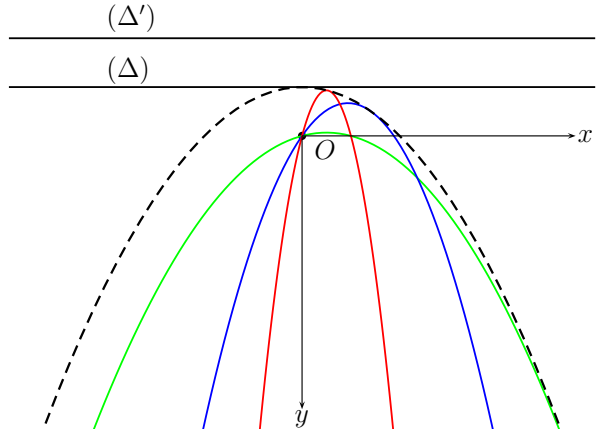}
\end{center}
\caption{\label{FigPara1} A few trajectories corresponding to various
shooting angles in a constant gravitational field, and their envelope (dotted
line).}
\end{figure}

Eliminating the time, $t$, in Eq.~(\ref{C:eq:para1}), and introducing the natural
length scale $a=v_0^2m/f$ of the problem leads to the well-known parabola
\begin{equation}\label{C:eq:para2}
y={x^2\over 2a\cos^2\alpha}+x \tan\alpha~.
\end{equation}
Examples are shown in Fig.~\ref{FigPara1}. Each
trajectory is drawn for both positive and negative times, $t=0$ corresponding
to the time of firing from $O$. This is equivalent to putting together the
trajectories corresponding to angles $\alpha$ and $\alpha+\pi$.

Equation (\ref{C:eq:para2}) can now be seen from a different view point:
given a point $M$ of coordinates $x$ and $y$, is there any possibility to
reach it with the gun? The answer is known: nearby points, or points located
downstream of the field can be reached twice, by a  straight shot
 or  a bell-like  trajectory.  Points located far away, or too
much upstream are, however, out of reach. The limit is the parabola
\begin{equation}\label{C:eq:par3}
y=-{a\over 2}+{x^2\over 2a}~,
\end{equation}
as seen, e.g., by writing (\ref{C:eq:para2}) as a second-order equation in
$\tan\alpha$ and requiring its discriminant to vanish. This envelope is shown in
Fig.~\ref{FigPara1}.
\subsection{Geometric solution}
The parabolas (\ref{C:eq:para2}) have in common a point $O$, and their
directrix, $\Delta$, which is located at $y=-a/2$. The equation can, indeed, be
read as
\begin{equation}\label{C:eq:par4}
(y+a/2)^2=(x-x_F)^2+(y-y_F)^2~,\quad x_F=-a\sin(2\alpha)/2,\
y_F=a\cos(2\alpha)/2~,
\end{equation}
revealing the focus $F$ located at $(x_F, y_F)$, i.e., on a circle, with centre at $O$, of radius
$a/2$, at an angle $\beta=2\alpha-\pi/2$ with the horizontal.

\begin{figure}[!!ht]
\begin{center}
\includegraphics{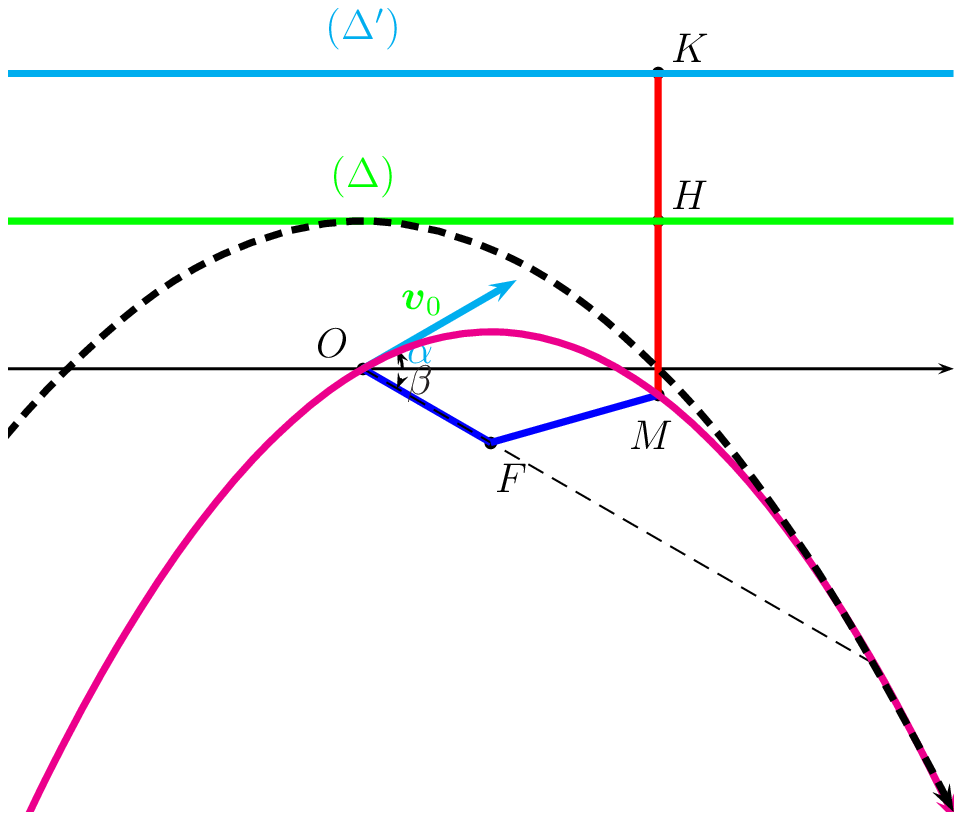}
\end{center}
\caption{\label{FigPara2} Geometrical construction of the envelope of the
ballistic parabolas.}
\end{figure}

The geometric construction follows, as shown in Fig.~\ref{FigPara2}.
A current point $M$ of a trajectory fulfills $MF=MH$, with the notation of
the figure. If $\Delta'$ is parallel to the common directrix $\Delta$, at
a distance $a/2$ further up, then the distance $MK$ to $\Delta'$ and the
distance $MO$ to the origin obey
\begin{equation}\label{proofpara}
MO\le MF+FO=MH+HK=MK~,
\end{equation}
this demonstrating
that the points within reach of the gun lie within a parabola of focus $O$ and
directrix $\Delta'$.
The equality is satisfied when $M$, $F$ and $O$ are aligned. From
$\beta=2\alpha-\pi/2$, a result pointed out by Macklin \cite{Macklin} is
recovered, that the tangent to the envelope is perpendicular to the initial
velocity of the trajectory that is touched. (If $M$ is on the envelope, $MO=MK$
and the tangent is the inner bisector of $\widehat{OMK}$.)
\subsection{A family of circles}
\begin{figure}[!!ht]\begin{center}
\includegraphics{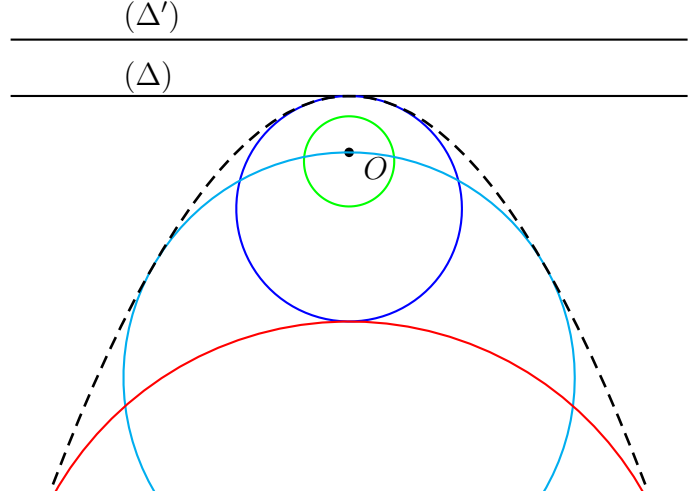}\end{center}
\caption{\label{FigPara3} The safety parabola can be seen as surrounding the
circles made at a given time $t$ by the projectiles shoot at once in all
directions with the same velocity. For small $t$, the circle does not touch the safety parabola.}
\end{figure}

A more peaceful view at the safe domain is that of an ideal firework:
projectiles of various angle $\alpha$ are fired all at once, with the same
velocity \cite{Donnelly}. At a given time $t$, they describe a circle (a sphere
in space)
\begin{equation}\label{C:eq:circ1}
x^2+(y-ft^2/(2m))^2=(v_0 t)^2~,
\end{equation}
with the centre at $\{0,ft^2/(2m)\}$, i.e., falling freely, and a growing radius
$v_0 t$. The problem of safety now consists of examining whether
Eq.~(\ref{C:eq:circ1}) has any solution in $t$ for given $x$ and $y$.  This
is a mere second-order equation, whose vanishing discriminant leads
back to the  parabola (\ref{C:eq:par3}). Figure \ref{FigPara3} show a few
circles whose envelope is this parabola.
%
\section{Coulomb or Kepler motion}\label{Kepler}
\subsection{Family of satellites}
\begin{figure}[!!ht]
\begin{center}
\includegraphics{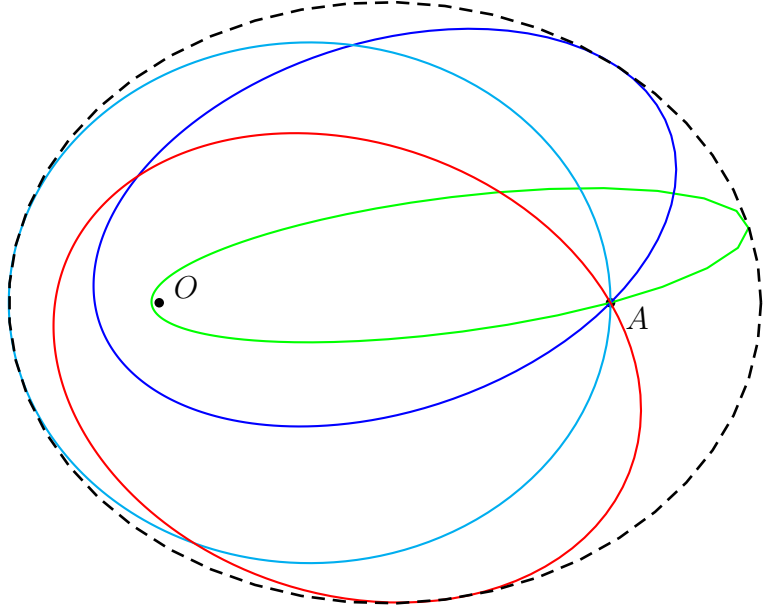}
\end{center}\caption{\label{FigEll1} Trajectories of satellites launched from $A$
 with the same velocity, but different initial direction. The envelope(dotted
line) is an ellipse of foci $O$ and $A$.}
\end{figure}

Let us consider an attractive Coulomb or Kepler potential $V=-K/r$, $K>0$,
centered
in $O$. If a particle of mass $m$ is fired from $A$ $(r_0, 0)$, with a velocity
$\vec{v}_0$, whose angle with $OA$ is $\alpha$, then the trajectory obeys the
equation \cite{Goldstein}
\begin{equation} \label{eq:Coul:1}
u''+u={m K\over \mathcal{L}^2}~,\quad u(0)={1\over r_0}~,\quad u'(0)=-{\cos\alpha\over r_0
\sin\alpha}~,
\end{equation}
where $\mathcal{L}=mr_0v_0\sin\alpha$ is the orbital momentum, which is
proportional to the constant areal velocity, and $u=1/r$, $u'={\rm d}u/{\rm
d}\theta$, etc. 
The solution is thus
\begin{equation} \label{eq:Coul:2}
u={K(1-\cos\theta)\over mr_0^2v_0^2\sin^2\alpha}+{\cos\theta\over r_0}
-{\sin\theta\cos\alpha\over r_0\sin\alpha}~,
\end{equation}
A few trajectories are shown in
Fig.~\ref{FigEll1}, together with their envelope, which is an ellipse with
foci, $O$, the centre of force, and $A$, the common starting point.
The envelope is easily derived by elementary calculus. Equation
(\ref{eq:Coul:2}), for a given point characterized by $u$ and $\theta$,
should have acceptable solutions in $\alpha$. This is a mere second order
equation in $\cot\alpha$, and the vanishing of its discriminant gives the
border of the safe domain.

\begin{figure}[!!hb]
\begin{center}
\includegraphics{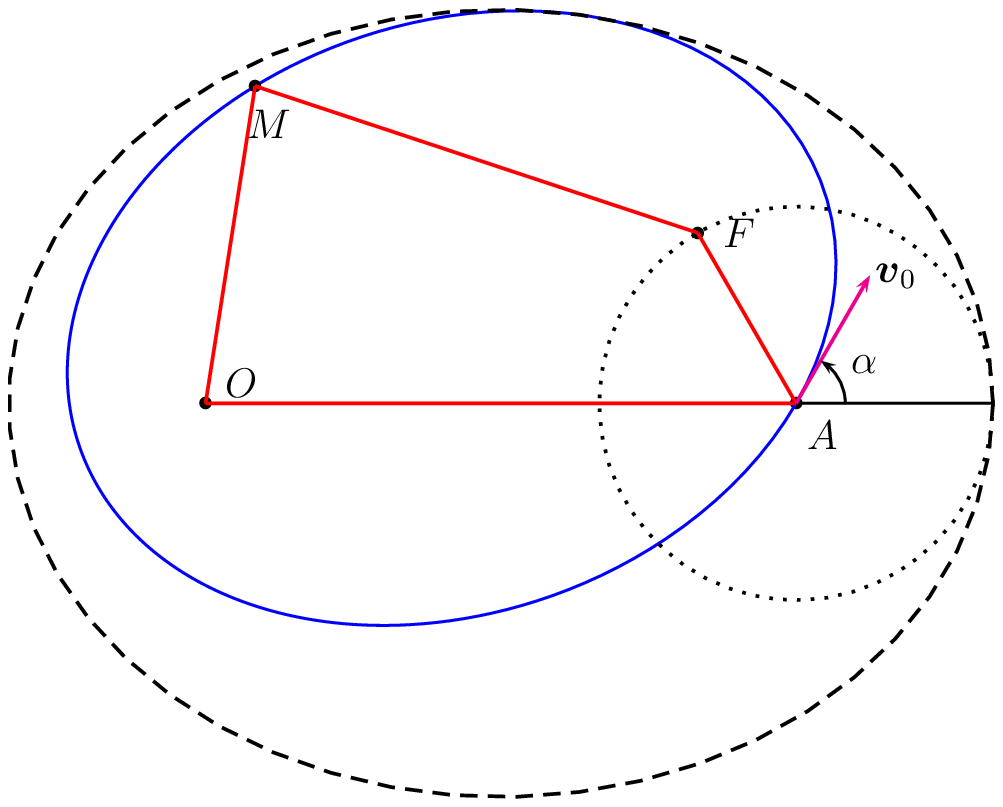}
\end{center}
\caption{\label{FigEll2} Geometric construction of the envelope of
trajectories of
satellites launched from $A$ with the the same velocity in various directions.}
\end{figure}

A geometric derivation of the envelope gives an answer even faster. All
trajectories have same energy, and hence the same axis $2a$, since $E=-K/(2a)$
\cite{Goldstein}.
Hence the second focus, $F$ is on a circle of centre $A$, and radius $2a-
r_0$. The initial velocity is one of the bisectors of $\widehat{OAF}$. For
any point $M$ on the trajectory, 
\begin{equation}\label{eq:Coul:3}
MO+MA\le MO+MF+FA=4a-r_0~,
\end{equation}
which proves the property. One further sees that the envelope is touched when
$M$, $F$ and $A$ are aligned. This is illustrated in Fig.~\ref{FigEll2}.
\subsection{Rutherford scattering}
\begin{figure}[!!ht]\begin{center}
\includegraphics{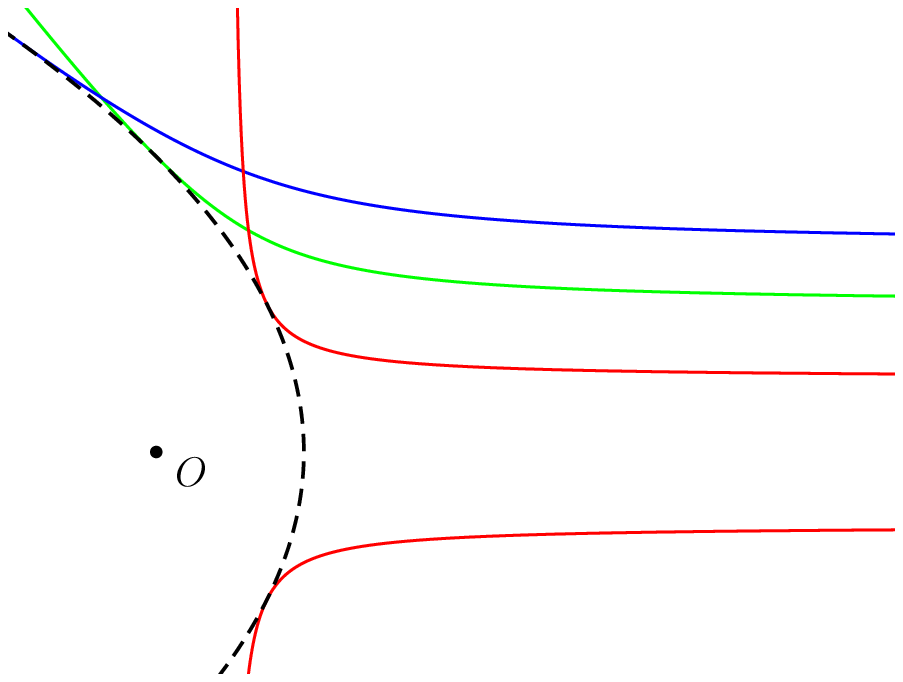}
\end{center}
\caption{\label{RuthInf}
Trajectories in a repulsive Coulomb potential with same velocity but 
varying impact parameter.}\end{figure}

As a variant, consider now the case of a repulsive interaction, $V=K/r$,
$K>0$, as in Rutherford's historical experiment.
The simplest case is that of particles sent from very far away with  the same
velocity $\vec{v}_0$ but different values of the impact parameter $b$, this
resulting in  varying orbital momenta $\mathcal{L}$. Examples are shown in
Fig.~\ref{RuthInf}. We have a family of hyperbolas
\begin{equation}\label{eq:RuthInf1}
u=-{a(1-\cos\theta)\over b^2} + {\sin\theta\over b}~,
\end{equation}
where $a=K/(mv_0^2)$. This second-order equation in
$b^{-1}$ has  real solutions if
\begin{equation}\label{eq:RuthInf2}
u\le{1+\cos\theta\over4a}~,
\end{equation}
corresponding to the outside of a parabola of focus $O$, also shown in
Fig.~\ref{RuthInf}.

\begin{figure}[!!ht]
\begin{center}
\includegraphics{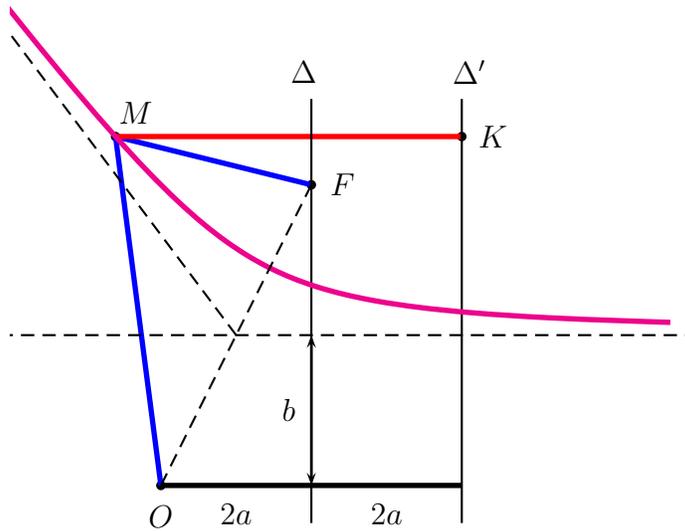}
\end{center}
\caption{\label{FigRuthGeo} Geometric proof that the Rutherford trajectories
with the same energy delimit a safe region.}
\end{figure}

The geometric interpretation is the following. All trajectories have the
same energy $E=mv_0^2/2$, and hence the same axis $2a$ since $E=K/(2a)$,
very much analogous to $E=-|K|/(2a)$ for ellipses in the case of attraction
and negative energy. Each hyperbola has a focus $O$, and second focus $F$ 
on the line $\Delta$, perpendicular to the initial asymptote at distance $2a$
from $O$. The middle of $OF$ lies on this asymptote, whose position is
determined by the impact parameter $b$. Let $\Delta'$ be parallel to
$\Delta$, at a further distance $2a$. If $M$ is on a trajectory, and is
projected on $\Delta'$ at $K$, then
\begin{equation}\label{eq:RuthInf3}
MK-MO\le MF+2a-MO=0~,
\end{equation}
with saturation when $M$, $F$ and $K$ are aligned. See Fig.~\ref{FigRuthGeo}.
\subsection{Scattering from finite distance}
\begin{figure}[!!ht]\begin{center}
\includegraphics{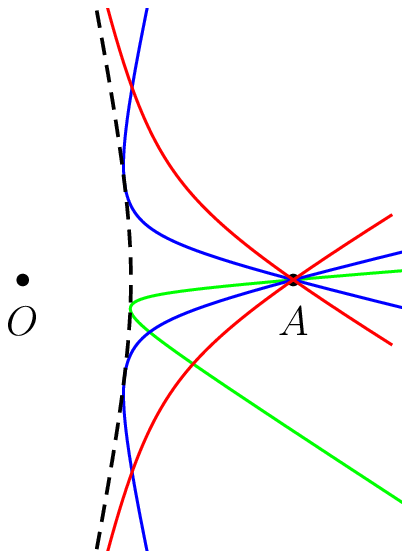}\ \includegraphics{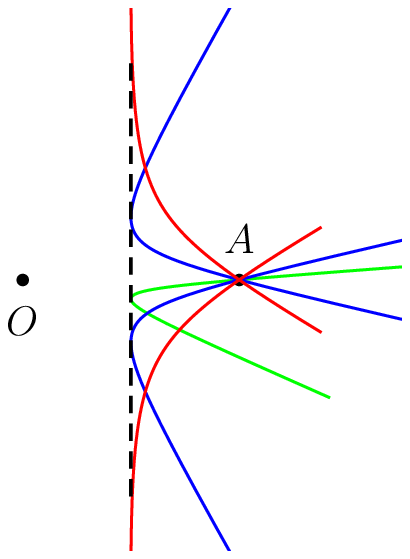}\ \includegraphics{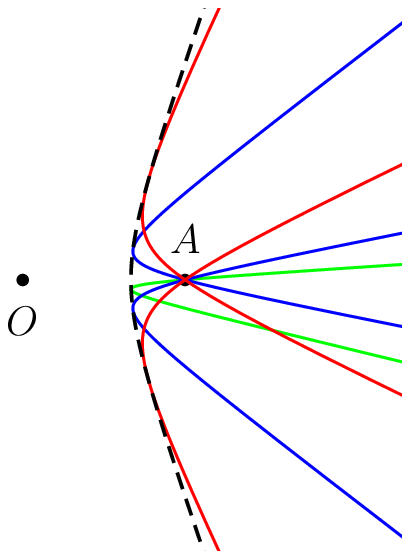}
\end{center}
\caption{Rutherford scattering from $r_0=5a/2$ (left), $r_{0}=2a$ (centre) and
$r_{0}=3a/2$ (right). The envelope is shown 
as a dotted line.\label{FigRuthFinite}}
\end{figure}

A simple generalization consists of considering particles launched in the
repulsive Coulomb field from a point $A$, at finite distance $r_{0}$ from the
centre  of force $O$.
The kinetic energy, written as $mv_0^2/2=K/(2a)$,
fixes the length scale $a$. The problem has been studied, e.g., by Samengo and
Barrachina \cite{Barrachina}, who discussed glory- and rainbow-like
effects. Some trajectories and their envelope and shown in
Fig.~\ref{FigRuthFinite}. It can be seen, and proved that
\begin{itemize}\itemsep-4pt
\item For $r_{0}>2a$, the envelope is a branch of hyperbola with $O$ as the
inner focus. In the limit $r_0 \to\infty$, we obtain the parabola of ordinary
Rutherford scattering.
\item For $r_{0}=2a$, the envelope is simply the mediatrix of $OA$.
\item For $r_{0}<2a$, the envelope is a branch of hyperbola with $O$ as the
external focus.
\end{itemize}
\section{Harmonic potential}\label{Harmo}
\subsection{Firing in various directions}
\begin{figure}[!!ht]\begin{center}
\includegraphics[width=.8\textwidth]{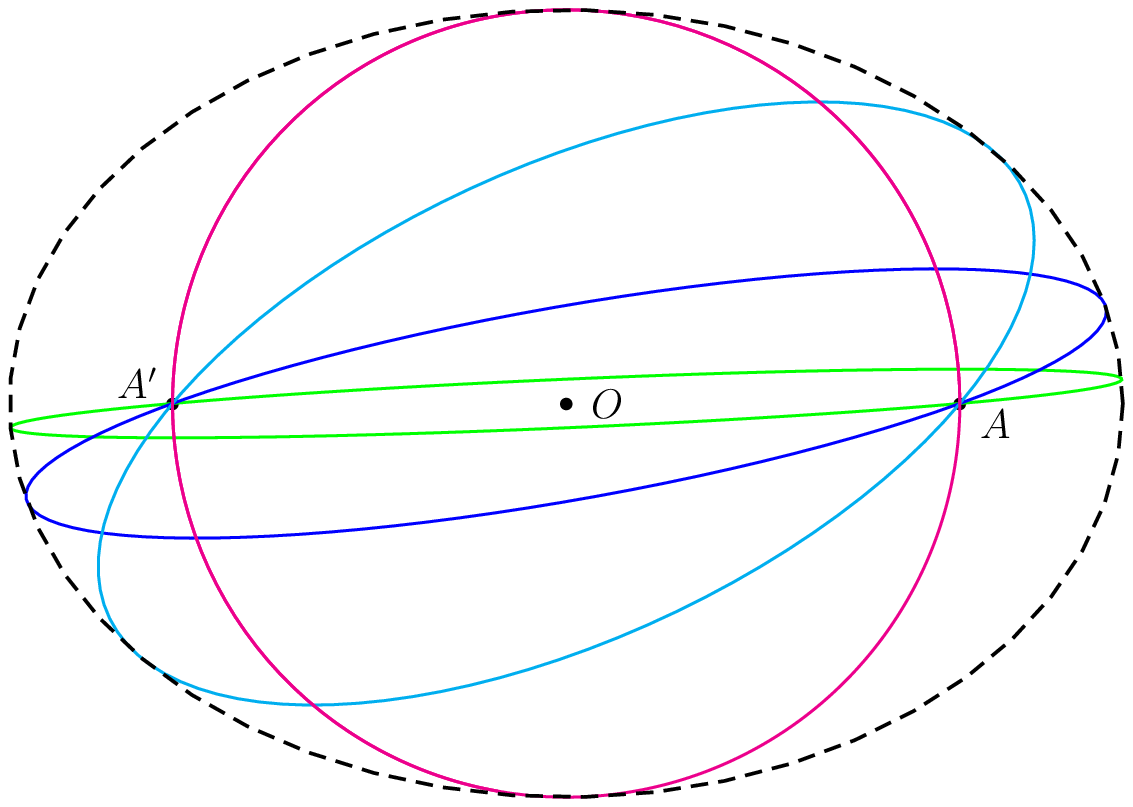}\end{center}
\caption{\label{FigHarmo1} Trajectories in a harmonic potential with 
varying angle for the initial velocity.}\end{figure}

We now consider a particle of mass $m$ in a potential $kr^2/2$. Figure \ref{FigHarmo1} shows a few
trajectories from $A$, located at $(x_0,0)$,
with an initial velocity $\vec{v}_0$ of given modulus and varying angle. 

This central-force problem is more easily solved directly in 
Cartesian coordinates, in contrast with most central forces problems,
for which the use of polar coordinates is almost mandatory. One obtains 
\begin{equation}\label{harmo:eq1}
x(t)=x_0\cos (\omega t)+ \ell_0 \cos \alpha\sin  (\omega t)~,\quad
y(t)=\ell_0\sin\alpha\sin(\omega t)~,
\end{equation}
where $\omega=\sqrt{k/m}$ and $\ell_0=v_0/\omega$, or, equivalently,  the algebraic equation 
\begin{equation}\label{harmo:eq1a}
\left({x-y\cot\alpha\over x_{0}}\right )^2+\left( {y\over
\ell_0 \sin\alpha}\right)^2=1~,
\end{equation}
which can be read as a second-order equation in $\cot\alpha$. The
condition to have real solutions for given $x$ and $y$ defines a domain
limited by the ellipse  
\begin{equation}\label{harmo:eq2}
{x^2\over x_0^2+\ell_0^2}+{y^2\over \ell_0^2}=1~,
\end{equation}
 with centre $O$ and foci $A$, and its symmetric $A'$. The points $A$ and $A'$
belong to all trajectories. This envelope is also shown in
Fig.~\ref{FigHarmo1}.

\subsection{Firing all projectiles simultaneously}
\begin{figure}[!!ht]\begin{center}
\includegraphics[width=.7\textwidth]{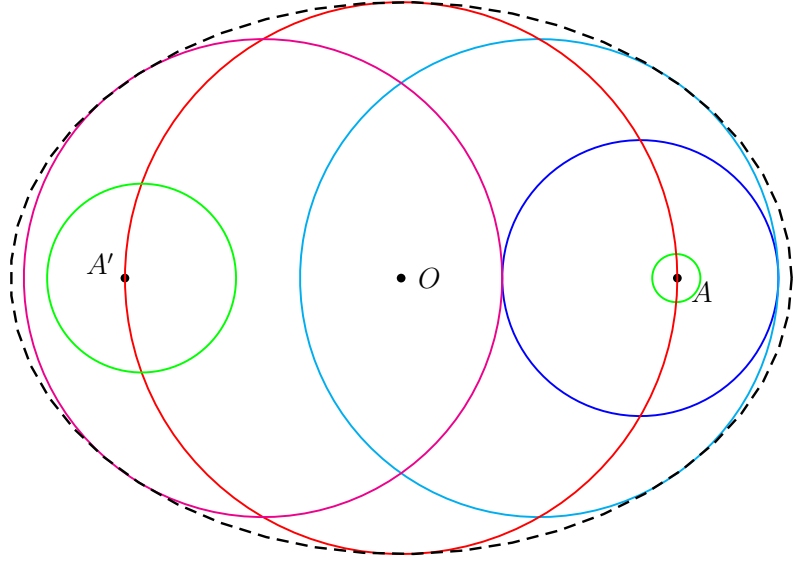}
\end{center}
\caption{\label{FigHarmo2}%
The projectiles, submitted to an harmonic potential centered at $O$, are 
all fired at $t=0$ with the same velocity, but in different directions. 
At any future time $t$, they describe a circle (a sphere in space) oscillating 
back and forth between $A$ and $A'$ with a radius of varying length.}
\end{figure}

The ``fireball'' view point leads to similar equations. If all
projectiles are fired all at once, they describe, in a plane, at time $t$,
the circle
\begin{equation}\label{harmo:eq3}
(x-x_0 \cos (\omega t))^2+y^2 = \ell_0^2\sin^2 (\omega t)~,
\end{equation}
whose radius $|\ell_{0}\sin(\omega t)|$ and centre position $x_{0}\cos (\omega t)$ oscillate. 
For given $x$ and $y$, this is, again, a second-order polynomial, now in
$\cos (\omega t)$, and  from its discriminant, the equation
(\ref{harmo:eq2}) of the envelope is recovered.
Figure \ref{FigHarmo2} shows the envelope surrounding the circles and touching
some of those.
\subsection{Geometric construction}
The geometric construction of this envelope can be carried out as follows. All
trajectories have again the same energy, $E=k(x_{0}^2+\ell_{0}^2)/2$, since only the angle of the initial
velocity is varied. This means that all ellipses have same quadratic sum
$a^2+b^2$ of semi-major and semi-minor axes. There exists, indeed, a basis,
where, after shifting time, the motion reads $X=a\cos(\omega t)$, $Y=b\sin(\omega t)$. 
If one recalculates the energy in this basis at $t=0$, one finds a potential term $ka^2/2$ and a kinetic term $m\omega^2 b^2/2$, and hence
 $E=k(a^2+b^2)/2$.

Now, if  $M$ a running point of a trajectory $T$
\begin{equation}\label{harmo:eq4}
MA+MA'\le \sup_{P\in T} (PA+PA')=2\sqrt{x_0^2+\ell_0^2}~,
\end{equation}
corresponding, indeed, to an ellipse of foci $A$ and $A'$.

A theorem is used here, that is not too well known, though it turns out
(after several investigations of the author) that it is at the level of next-to-elementary
geometry. It is described below.
\subsection{A theorem on ellipses}
\noindent\textbf{Theorem:} {\sl If $A\in T$ and $A'\in T$ form a diameter of an
ellipse $T$, and $M$ denotes a running point of $T$, the maximum of the sum of
distances
\begin{equation}\label{harmo:eq5}
\sup_{M\in T}(MA+MA')~,
\end{equation}
is independent of $A$, with value $2\sqrt{a^2+b^2}$, where $a$ and $b$
denote the semi transverse  and conjugate axes of $T$.}

The proof can be found in Ref.~\cite[p.~350]{Berger}.
It is linked to a consequence of the Poncelet theorem formulated by 
Chasles.

\begin{figure}[!!ht]\begin{center}
\includegraphics[width=.6\textwidth]{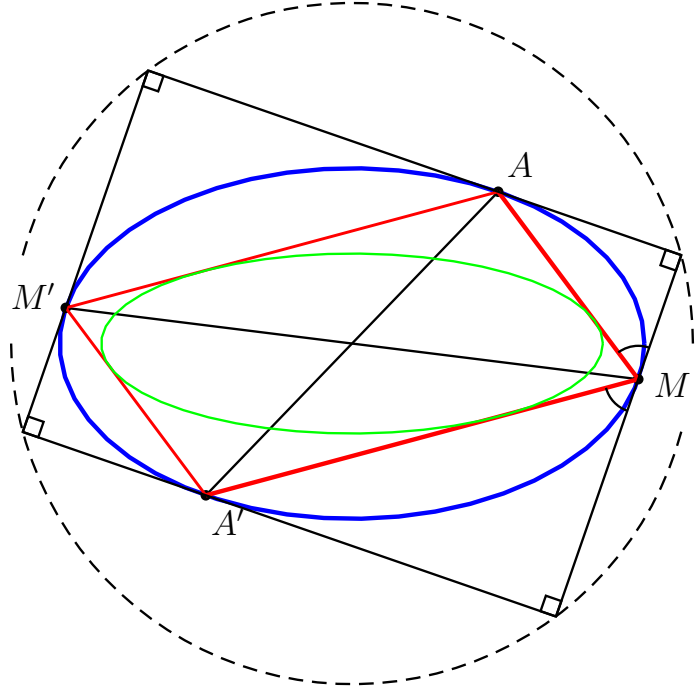}\end{center}
\caption{\label{FigHarmo3}
An ellipse of semi-axes $a$ and $b$, a
 diameter $AA'$ and the points $M$ and $M'$ such as $MA+MA'$ is maximal. 
The result is independent of $A$. The tangents in $A$ and $M$ are orthogonal, 
and thus intersect on the orthoptic curve of the ellipse, which is a circle.
The sides of the parallelogram are tangent to a homofocal ellipse of semi-axes
$a^2/\sqrt{a^2+b^2}$ and $b^2/\sqrt{a^2+b^2}$.}
\end{figure}

Steps in understanding the above property include:
\begin{itemize}\itemsep-2pt
\item The maximum is reached twice, for say, $M$ and $M'$,  $AMA'M'$ forming a
parallelogram, as shown in Fig.~\ref{FigHarmo3}.
\item By first order variation, the tangent in $M$ is a bissectrix of
$\widehat{AMA'}$.
\item The tangent in $M$ is perpendicular to the tangent in $A$. This
provides the non-trivial result that if $M$ maximizes $MA+MA'$, conversely,
$A$ (or $A'$) 
maximizes the sum of distances to $M$ and $M'$.
\item The tangents in $A$, $M$, $A'$, and $M'$ forming a rectangle, the Monge
theorem \cite[p.~332]{Berger} applies, stating that the orthoptic curve of the
ellipse (set of points from which the ellipse is seen at 90$^\circ$), is a circle
of radius $\sqrt{a^2+b^2}$, see Fig.~\ref{FigHarmo3}.
\item The sides of the parallelogram $AMA'M'$ are tangent to an ellipse $T'$
with same foci as $T$, but flattened, with semi-axes 
\begin{equation}\label{harmo:eq6}
a'={a^2\over \sqrt{a^2+b^2}}~,\quad
b'={a^2\over \sqrt{a^2+b^2}}~.
\end{equation}
\end{itemize}

Note that if one tries to demonstrate this theorem by straightforward
calculus, one generally writes down cumbersome equations. One of the simplest
-- if not the simplest -- methods, would consist of starting from the 
trajectories (\ref{harmo:eq1}), identifying there the most general set of
ellipses of given $a^2+b^2$, and calculating the envelope (\ref{harmo:eq2}),
which is easily identified as an ellipse of foci $A$ and $A'$ and major axis
$2\sqrt{a^2+b^2}$. It thus follows that on each trajectory,
$MA+MA'\le2\sqrt{a^2+b^2}$,
with saturation when the trajectory touches its envelope. 
\section{Conclusions}
Conic sections are encountered in classical optics, where they provide a design for ideal mirrors with perfect refocusing properties of light rays emitted by a suitably-located point-source.

Trajectories in elementary mechanics simple potentials such as a linear, Coulomb or quadratic potential, also follow conic sections. When the angle of the initial velocity is varied, one gets a family of trajectories with the same total energy. Their envelope sets the limits of the safe domain. This envelope can be deduced by recollecting astute though somewhat old-fashioned methods of geometry courses.

If the potential becomes more complicated or if one uses relativistic kinematics, the envelope has to be derived by calculus, but the techniques can be probed first on these cases where a purely geometric solution is available.

\subsection*{Acknowledgements} Useful information from X.~Artru,
J.-P.~Bourguignon, A.~Connes and M.~Berger, and comments by A.J.~Cole are
gratefully acknowledged.


\begin{thebibliography}{9}
\def\AJP{Am. J. Phys.}
%
\bibitem{AJP-EJP} See, for instance, 
A. Gonz\'{a}lez-Villanueva, E. Guillaum\'{i}n-Espa\~{n}a,
R.P. Mart\'{i}nez-y-Romero, H.N. N\'{u}\~{n}ez-Y\'{e}pez and A. L. Salas-Brito,
Eur.\ J.\ Phys. \textbf{19}, 431 (1998);
S.K. Bose, \AJP\ \textbf{53}, 175 (1985);
D.~Derbes, \AJP\ \textbf{69}, 481 (2001); 
Th.A. Apostolatos, \AJP\ \textbf{71} 261 (2003); 
D.M. Williams, \AJP\ \textbf{71}, 1198 (2003); and references therein.
%
\bibitem{Donnelly} D.~Donnelly, \AJP\ \textbf{60}, 1149 (1992).
%
\bibitem{French}A.P.~French, \AJP\ \textbf{61}, 805  (1993).
%
\bibitem{Warner} R.E.~Warner and L.A.~Huttar, \AJP\ \textbf{59}, 755 (1991).
%
\bibitem{Macklin}Ph.A.~Macklin, \AJP\ \textbf{55}, 947 (1987).
%
\bibitem{Barrachina}I.~Samengo and R.O.~Barrachina,
Eur.\ J.\ Phys. \textbf{15}, 300 (1994).
%
\bibitem{Goldstein} See, for instance, H.~Goldstein, Ch.~Poole and J.~Safko,
\textsl{Classical Mechanics}, 3rd ed. (Addison-Wesley, New-York, 2002).
%
\bibitem{Berger}M.~Berger, {\it G\'eom\'etrie}, Tome 2 (Nathan, Paris, 1990).
\end{thebibliography}
\end{document}